\documentclass{article}

\PassOptionsToPackage{numbers, compress}{natbib}
\usepackage[preprint]{neurips_2023}

\usepackage[utf8]{inputenc} 
\usepackage[T1]{fontenc}    
\usepackage{hyperref}       
\usepackage{url}            
\usepackage{booktabs}       
\usepackage{amsfonts}       
\usepackage{nicefrac}       
\usepackage{microtype}      
\usepackage{xcolor}         
\usepackage{amsmath}
\usepackage{amssymb}
\usepackage{mathtools}
\usepackage{amsthm}
\usepackage{multirow}
\usepackage{wrapfig}
\usepackage[capitalize,noabbrev]{cleveref}
\usepackage{graphicx}
\usepackage{subfigure} 
\usepackage{float}
\usepackage{algorithm} 
\usepackage{algorithmic}
\usepackage{textcomp} 
\usepackage{listings} 
\usepackage{enumitem}
\usepackage{utfsym}
\usepackage{times}
\usepackage{latexsym}
\usepackage{fancyvrb}
\usepackage{CJKutf8}

\newcommand{\commentout}[1]{}

\title{FireRedTTS-1S: An Upgraded Streamable Foundation Text-to-Speech System}

\author{
FireRed Team\thanks{Authors (alphabetical order): Hao-Han Guo, Yao Hu, Fei-Yu Shen, Xu Tang, Yi-Chen Wu, Feng-Long Xie, Kun Xie. Corresponding Author: Fenglong Xie (fenglongxie@xiaohongshu.com)} \\
Xiaohongshu \\
}

\begin{document}
\maketitle

\begin{abstract}

In this work, we upgrade FireRedTTS to a new version, FireRedTTS-1S, a high-quality streaming foundation text-to-speech system. FireRedTTS-1S achieves streaming speech generation via two steps: text-to-semantic decoding and semantic-to-acoustic decoding. In text-to-semantic decoding, a semantic-aware speech tokenizer converts the speech signal into semantic tokens, which can be synthesized from the text via a language model in an auto-regressive manner. Meanwhile, the semantic-to-acoustic decoding module simultaneously translates generated semantic tokens into the speech signal in a streaming way. We implement two approaches to achieve this module: 1) a chunk-wise streamable flow-matching approach, and 2) a multi-stream language model-based approach. They both present high-quality and streamable speech generation but differ in real-time factor (RTF) and latency. Specifically, flow-matching decoding can generate speech by chunks, presenting a lower RTF of 0.1 but a higher latency of 300ms. Instead, the multi-stream language model generates speech by frames in an autoregressive manner, presenting a higher RTF of 0.3 but a low latency of 150ms. In experiments on zero-shot voice cloning, the objective results validate FireRedTTS-1S as a high-quality foundation model with comparable intelligibility and speaker similarity over industrial baseline systems. Furthermore, the subjective score of FireRedTTS-1S highlights its impressive synthesis performance, achieving comparable quality to the ground-truth recordings. These results validate FireRedTTS-1S as a high-quality streaming foundation TTS system.

\end{abstract}

\section{Introduction}
\label{sec:intro}

Under the guidance of scaling law, language modeling \cite{xie2024towards, chen2024next} has been validated as a promising solution for text-to-speech synthesis (TTS), such as BASE-TTS \cite{lajszczak2024base}, Seed-TTS \cite{anastassiou2024seed}, CosyVoice \cite{du2024cosyvoice, du2024cosyvoice2}, TouchTTS \cite{song2024touchtts}, IndexTTS \cite{deng2025indextts}, Spark-TTS \cite{wang2025spark}, and FireRedTTS \cite{guo2024fireredtts} proposed by us. These systems can produce high-quality speech with high similarity to the target voice in terms of voice quality and speaking style, given only one reference audio.

This impressive achievement attracts much more attention from the industry and academia. The wide use of this technology also puts higher requirements on foundation TTS: 1) higher synthesis quality: the system should synthesize intelligible and natural speech with arbitrary voices with various characteristics and styles; 2) real-time generation: the system should output the speech signal instantly once receiving the text. It makes FireRedTTS, as an industry-level foundation model proposed one year ago, difficult to meet these more strict requirements, calling for an upgrade of this system.

To tackle this challenge, in this work, we upgrade FireRedTTS \cite{guo2024fireredtts} (also denoted by FireRedTTS-1) to a new version, FireRedTTS-1S, a streaming foundation TTS system. We achieve a high-quality streaming synthesis process via two modules, text-to-semantic decoding and semantic-to-acoustic decoding. Text-to-semantic decoding employs a language model to convert the text to semantic tokens of speech auto-regressively. In semantic-to-acoustic decoding, we upgrade both flow-matching and language-model-based approaches proposed in FireRedTTS-1 to simultaneously generate the speech signal from the semantic tokens in a streaming way. First, the upgraded flow-matching approach can generate speech by chunks, achieving streamable generation with a lower real-time factor (RTF). Meanwhile, we incorporate a multi-stream language model with a causal acoustic codec to generate speech by frames in an autoregressive manner, presenting a lower latency. Moreover, we scale up the dataset with around 500k hours of Chinese-English speech data to train this system. Finally, it performs higher-quality zero-shot voice cloning via in-context learning while achieving real-time streamable generation.

We conduct both objective and subjective tests to evaluate the synthesis performance of the proposed streaming foundation TTS system in the zero-shot (one-shot) setting. First, the result of the objective test validates that FireRedTTS-1S achieves significant improvement in intelligibility and speaker similarity over the previous version, presented as comparable to other SOTA streamable and non-streamable foundation TTS systems. Both semantic-to-acoustic approaches present similarly impressive performance. Moreover, the subjective test further demonstrates the impressive synthesis performance of FireRedTTS-1S, which not only outperforms the previous version significantly, but also presents human-level performance by a comparable subjective score to ground-truth recordings. Combined with the streamable generation capability, this system is validated as a high-quality streaming foundation TTS system. The inference code of the foundation model and the demo page are available at \url{https://fireredteam.github.io/demos/firered_tts_1s}.

\section{FireRedTTS-1S}

\begin{figure}[htp]
\centering
\includegraphics[width=\linewidth]{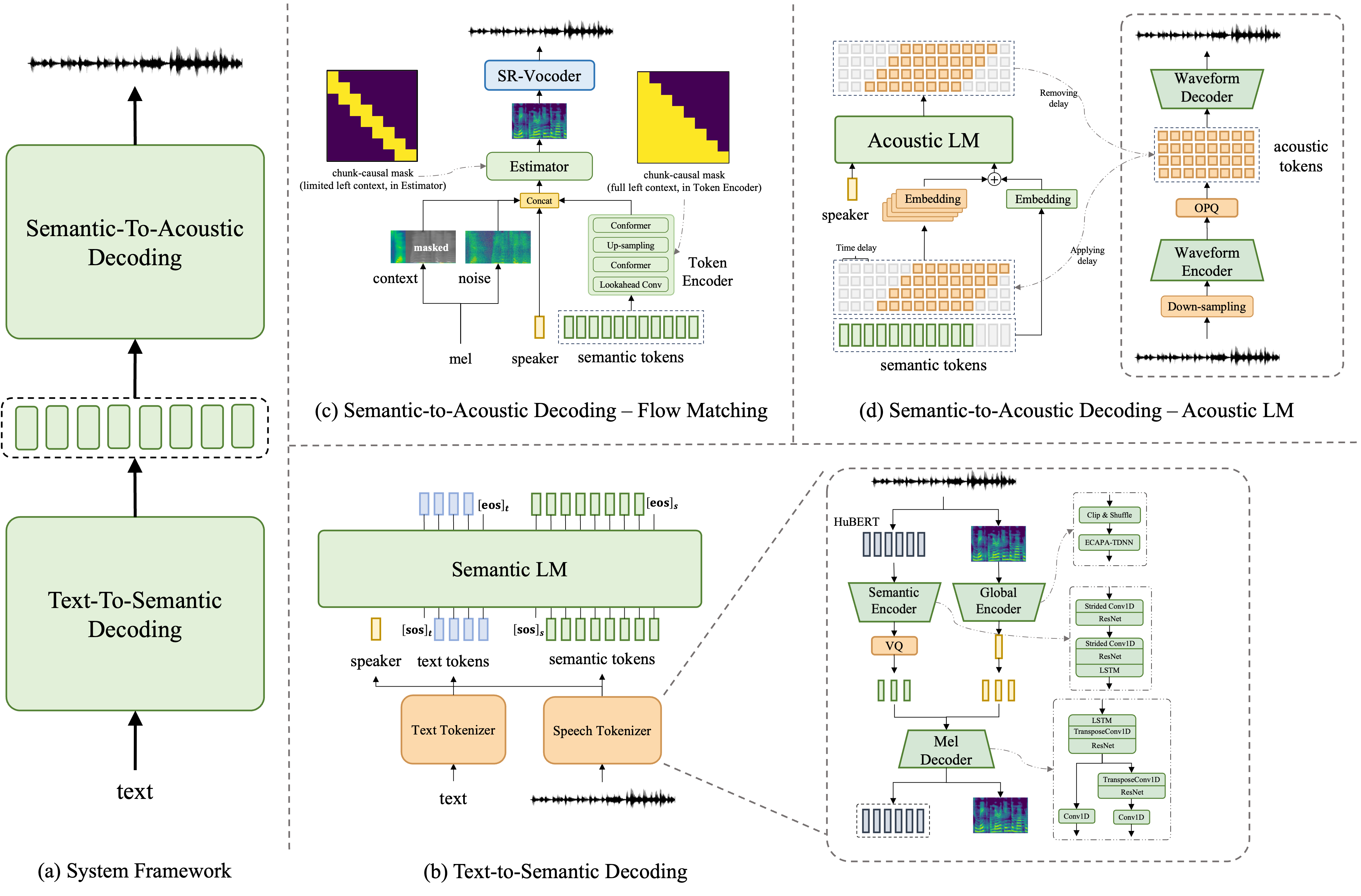}
\caption{An overview of the FireRedTTS-1S: In text-to-semantic decoding, a speech tokenizer is trained to extract semantic tokens and speaker embedding from the speech audio, and a semantic language model generates semantic tokens from the input text sequence and the speaker embedding from the reference audio. In semantic-to-acoustic decoding, we implement two approaches to translate semantic tokens into the speech waveform in a streaming way. In (c), we empower the flow-matching approach with the streamable decoding capability. In (d), a multi-stream acoustic LM with a ``delay pattern" is trained to map semantic tokens into the multi-stream acoustic token sequence, which is encoded from the speech signal and decoded back by the acoustic codec.}
\label{img:tts_model_framework}
\end{figure}

As shown in Figure \ref{img:tts_model_framework}, we propose a language-model-based streaming foundation TTS system. This system is composed of two modules: text-to-semantic decoding, and semantic-to-acoustic decoding.

\subsection{Text-to-Semantic Decoding}

First, the text-to-semantic decoding module aims to generate the discrete speech sequence mainly representing the semantic information of speech from the input text and the reference audio.

\subsubsection{Semantic-Aware Speech Tokenizer}

We still use the speech tokenizer proposed in FireRedTTS-1 \cite{guo2024fireredtts} to extract semantic tokens and speaker embeddings. The semantic encoder employs a pre-trained self-supervised learning (SSL) model, HuBERT \cite{hsu2021hubert}, to convert speech signals into a semantic embedding sequence, and then process it with a stack of convolutional layers and an optimized vector-quantization \cite{guo2024addressing} layer into a token sequence with a frameshift of 40ms and a codebook size of 16,384 tokens. Meanwhile, the acoustic encoder, i.e. an ECAPA-TDNN module \cite{dawalatabad2021ecapa}, maps the Mel spectrogram of the input audio into a global embedding representing speaker information. The pre-processing strategy, "Clip\&Shuffle", is applied to the Mel spectrogram to remove short-time variant information. It randomly selects a segment with 25\% to 75\% lengths from Mel spectrogram, and shuffles it by clipping the segment into 1-second slices. Finally, the decoder reconstructing SSL features and acoustic features is only used in training to guide two encoders to learn expected representations.

\subsubsection{Semantic LM}

As shown in Figure \ref{img:tts_model_framework} (a), we still formulate TTS as a next-token prediction task using a 30-layer decoder-only autoregressive Transformer (400M parameters) trained with a larger speech dataset with 500,000 hours of Chinese-English speech data collected by the protocol proposed in FireRedTTS-1. We process the text with the BPE-based text tokenizer \cite{whisper}, and embed the prompt audio into the speaker embedding by the speech tokenizer. The speaker embedding, text sequence, and speech sequence are embedded respectively and then concatenated together for the next-token prediction training. At inference, we employ the in-context learning (ICL) strategies. We extract the transcribed text, semantic tokens, and speaker embedding from the reference audio. Then, we concatenate the speaker embedding, transcribed text, input text, and semantic tokens together to the semantic LM for TTS synthesis. This approach can produce speech with a consistent speaking style and voice quality to the prompt audio, widely used in zero-shot or one-shot voice cloning.

\subsection{Semantic-to-Acoustic Decoding}

The semantic-to-acoustic decoding module aims to translate semantic tokens into the corresponding speech signal in a streaming way. We implement two approaches to achieve this module: a streamable flow-matching approach and a multi-stream language-model-based approach. 

\subsubsection{Streamble Flow-Matching}
Flow matching \cite{lipman2022flow} has been validated as an effective approach for high-quality speech synthesis. It has been widely incorporated with neural vocoders \cite{hifigan, bigvgan} in foundation TTS systems \cite{anastassiou2024seed, du2024cosyvoice, du2024cosyvoice2} to decode semantic tokens to the waveform. In this work, we upgrade the FireRedTTS-1 flow-matching model to achieve streamable, high-quality speech generation. In this approach, a streamable flow-matching model converts semantic tokens into the Mel spectrogram with the frequency band between 0 and 8kHz, then a BigVGAN-based \cite{bigvgan} super-resolution vocoder with pseudo-streaming decodes the spectrogram into the waveform with the sampling rate of 24kHz.

\textbf{Chunk-Wise Streaming Flow Matching:} The flow matching module comprises two key components: the token encoder and the estimator. Semantic tokens are first embedded and processed by a look-ahead convolutional layer that encodes the future 3-token context to enable smoother streaming chunk transitions. Then, they are further processed by two Conformer blocks with an intermediate upsampling layer in between for temporal alignment with the Mel spectrogram. 
The estimator employs a DiT-based \cite{peebles2023scalable} model architecture, and is enhanced with a two-layer causal convolutional encoder after each self-attention block. To achieve higher-quality voice cloning, we employ in-context learning by providing the initial 0-30\% of ground-truth Mel spectrogram as a context feature. The estimator's input concatenates the encoded token features, speaker embedding, context feature, and noisy mel spectrogram along the feature dimension to predict the flow-matching velocity field.
During training, we refer to CosyVoice2 \cite{du2024cosyvoice2} to randomly select from four variants of attention masks per sample for both the token encoder and estimator, i.e. (1) full attention, (2) fully causal, (3) 1-second chunk causal masks and (4) 2-second chunk causal masks. We notice that, at inference, the iterative denoising of flow matching needs to store intermediate key-value caches in transformer layers, leading to substantial GPU memory consumption. To mitigate this issue, we limit the left context to 2 seconds during training when employing chunk-wise causal masks. Finally, we implement a streamable flow-matching model with 150M parameters. It is trained in two stages: (1) 800k updates using only full attention masks, followed by (2) 100k updates with random selection among the four mask variants. This yields a unified model capable of both streaming and non-streaming inference.

\begin{figure}[htp]
\centering
\includegraphics[width=0.5\linewidth]{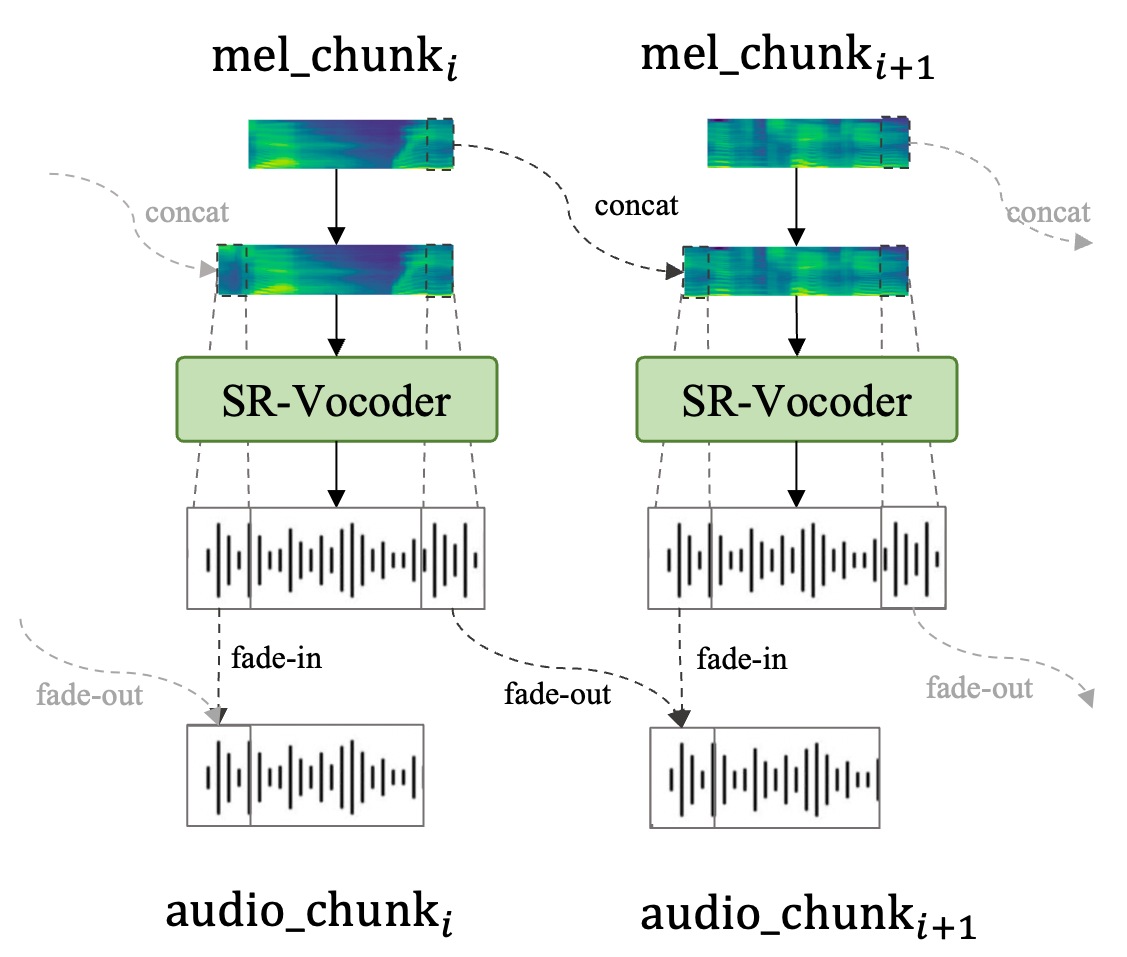}
\caption{The pseudo-streaming strategy for the neural vocoder.}
\label{img:vocoder}
\end{figure}


\textbf{Pseudo-Streaming BigVGAN Vocoding:} To achieve streaming vocoding for the generated Mel spectrogram, we apply a pseudo-streaming approach. As presented in Fig. \ref{img:vocoder}, each chunk of the generated Mel spectrogram $mel\_chunk_{i+1}$ is concatenated with its previous eight frames as the input of the neural vocoder. This approach can provide sufficient context to the first frames of the chunk for stable waveform generation. Then, the generated waveform for this chunk is overlapped with the generated waveform corresponding to the previous eight frames by fade-in/fade-out crossfading to reduce concatenation artifacts. 

The flow-matching approach can achieve both streaming and non-streaming inference. Non-streaming generation can process all semantic tokens in a parallelized way, presenting a higher throughput rate (lower RTF) but also higher latency. By chunk-wise generation, we achieve streaming generation by sacrificing the throughput rate for lower RTF. This approach can effectively control the balance between these two terms to better adapt to real-world applications. The theoretical latency bound is:
$$\text{latency}\leq (l+3) * t_s + t_a + t_c$$
where $l$ denotes the token count per chunk, with $t_s$, $t_a$, $t_c$ representing the average processing times for the semantic LM (per step), flow model, and BigVGAN, respectively. In our work, using a 1-second chunk size ($l=25$), the optimized system achieves real-time speech generation with the RTF of 0.1 and latency $< 300ms$ on an NVIDIA L20 GPU.

\subsubsection{Acoustic LM}

This approach comprises two models, first employing a multi-stream acoustic language model to convert semantic tokens into acoustic tokens with richer acoustic information, which are then generated into the corresponding speech signal by a causal acoustic codec simultaneously.

\textbf{Super-Resolution Causal Acoustic Codec:} First, we train an acoustic codec to convert the speech signal into acoustic tokens while reconstructing them back in a streamable way. The acoustic codec is implemented with a similar architecture to BigCodec \cite{xin2024bigcodec}, a BigVGAN-based \cite{bigvgan} CNN-based codec. Differently, we transform it into a streamable model by replacing all convolutional layers with causal convolutional layers and removing the upsampling/downsampling operations \cite{wu2024ts3} at the snake activation function. The vanilla VQ layer is replaced with the ordered product quantization (OPQ) proposed in SoCodec \cite{socodec} to learn an ordered multi-stream acoustic representation to adapt the multi-stream language model with a ``delay pattern" better. The codec receives the waveform with a lower sampling rate of 16kHz as the input, but outputs the waveform with a higher sampling rate of 24kHz. This super-resolution operation allows us to adapt the codec to the low-sampling-rate audio, which is easier to obtain in LM training.

In this work, the acoustic codec encodes the speech audio into a multi-stream acoustic sequence with eight codebooks, each consisting of 16,384 codewords, and a frameshift of 40ms. The model is trained with a batch size of 192 seconds and a constant learning rate of 1e-4. The training is conducted in two stages: 1) updating all parameters for 1 million iterations with the enabled stream-wise dropout of OPQ, and 2) only updating the decoder for 300,000 iterations with the disabled stream-wise dropout to achieve higher reconstruction quality.

\textbf{Acoustic LM:} Similar to the streamable decoder proposed in FireRedTTS, we train a multi-stream acoustic LM to generate acoustic tokens auto-regressively from semantic tokens and the speaker embedding. We apply a "delay pattern" \cite{copet2024simple} to the multi-stream acoustic sequence, i.e. applying each row with a delay of 1 step over the previous row, to achieve autoregressive generation along the time and the stream axis. To simultaneously convert the received semantic tokens from the semantic LM into the acoustic tokens for streaming generation, we propose merging them into one embedding sequence after processing them with different embedding layers. We apply a ``time delay'' of $d$ when combining these two sequences, i.e. i.e. $s_{i-d} + a_{i}$, where $a_{i}$ is the $i$-th acoustic embedding, $s_{i-d}$ is the $(i-d)$-th semantic embedding. This operation helps better predict acoustic tokens by providing more future information, but also causes more delay in waiting for sufficient semantic tokens to start generation. Finally, this embedding sequence is fed to a stack of transformer layers to generate the next-step acoustic tokens, and fed to the codec decoder to generate the corresponding waveform after removing the ``delay pattern".

At TTS inference, given the input text, the semantic LM starts to generate semantic tokens. The semantic-to-acoustic decoding starts simultaneously after receiving sufficient semantic tokens. The theoretical latency of FireRedTTS-1S can be calculated as follows:
\begin{equation}
    \text{latency} <= (d - 1) * t_{s} + m * (t_{s} + t_{a}) + t_{c}
\end{equation}
where $t_{s}$ is the average time running one step of the semantic LM, $t_{a}$ is the average time running one step of the acoustic LM, and $t_{c}$ is the average time decoding one frame of acoustic tokens. In our work, we select $d=8$ and $m=8$. The acoustic LM has 24 layers of 1536-dim causal transformer layers, and is trained for 200,000 iterations with a batch size of 6400 seconds. Finally, we achieve a real-time speech generation with a higher RTF of 0.3 but a latency $<150ms$ on L20 after engineering optimization.

\section{Results}

\subsection{Objective Evaluation}
\label{ssec:obj}

First, we conduct an objective evaluation of zero-shot voice cloning, by following the approach\footnote{The test sets and evaluation tools are available at \url{https://github.com/BytedanceSpeech/seed-tts-eval}} used in Seed-TTS \cite{anastassiou2024seed} and CosyVoice \cite{du2024cosyvoice}, to investigate the performance of FireRedTTS-1S as a foundation TTS system. This evaluation comprises three test sets, ``Test-ZH" with 2020 Chinese utterances from DiDiSpeech \cite{guo2021didispeech}, ``Test-EN" with 1088 English utterances from Common Voice \cite{ardila2019common}, and ``Test-ZH-Hard" with 400 complex Chinese utterances. The character/word error rate (CER/WER) and cosine similarity between speaker embeddings (SIM) are calculated with automatic speech recognition (ASR) \cite{gao2023funasr} and speaker recognition (SR) \cite{chen2022wavlm} models to measure the intelligibility and speaker similarity of the synthesized speech. Since each test case has only one reference audio for zero-shot voice cloning, we adopt full-ICL for FireRedTTS-1S.

As shown in Table \ref{tab:obj}, FireRedTTS-1S-FM and FireRedTTS-1S-LM achieve a similar performance on intelligibility and speaker similarity across all test sets, which are significantly better than the previous version. Compared with other streamable and non-streamable systems, FireRedTTS-1S also presents a strong competitiveness. First, in ``Test-ZH", it achieves the lowest CER and comparable SIM score over other systems, where FireRedTTS-1S-FM performs slightly better. Moreover, FireRedTTS-1S presents a more prominent improvement in the ``Test-EN" test set, presented as a better streaming TTS system over CosyVoice2. Even in the complex test set ``Test-Zh-Hard", where all systems exhibited a certain degree of degradation, FireRedTTS-1S still achieves relatively high performance, where FireRedTTS-1S-LM is only slightly worse than the non-streamable closed-source system, Seed-TTS.

\begin{table}[htp]
\centering
\begin{tabular}{c|c|cc|cc|cc}
\toprule
\multirow{2}{*}{\textbf{System}} & \multirow{2}{*}{\textbf{Streaming}} & \multicolumn{2}{c|}{\textbf{Test-ZH}} & \multicolumn{2}{c|}{\textbf{Test-EN}} & \multicolumn{2}{c}{\textbf{Test-ZH-Hard}} \\ \cmidrule{3-8}
 & & \multicolumn{1}{c}{\textbf{CER $\downarrow$}} & \multicolumn{1}{c|}{\textbf{SIM $\uparrow$}} & \multicolumn{1}{c}{\textbf{WER $\downarrow$}} & \multicolumn{1}{c|}{\textbf{SIM $\uparrow$}} & \multicolumn{1}{c}{\textbf{CER $\downarrow$}} & \multicolumn{1}{c}{\textbf{SIM $\uparrow$}} \\ \midrule
Human & - & 1.26 & 0.755 & 2.14 & 0.734 & - & - \\ \midrule
*Seed-TTS \cite{anastassiou2024seed}      & $\times$ & \textcolor{orange}{1.12} & \textcolor{green}{0.796} & 2.25 & \textcolor{green}{0.762} & \textcolor{green}{7.59} & \textcolor{green}{0.776} \\
MaskGCT \cite{wang2024maskgct}       & $\times$ & 2.27 & \textcolor{blue}{0.774} & 2.62 & \textcolor{blue}{0.714} & 10.27 & \textcolor{blue}{0.748} \\
F5-TTS \cite{chen2024f5}       & $\times$ & 1.56 & 0.741 & \textcolor{green}{1.83} & 0.647 & 8.67 & 0.713 \\
CosyVoice 2 \cite{du2024cosyvoice2}   & $\checkmark$ & 1.45 & \textcolor{orange}{0.753} & 2.38 & 0.654 & 8.08 & 0.732 \\ \midrule
FireRedTTS-1 \cite{guo2024fireredtts}  & $\times$ & 1.51 & 0.635 & 3.82 & 0.460 & 17.45 & 0.621 \\
FireRedTTS-1S-FM & $\checkmark$ & \textcolor{green}{1.00} & \textcolor{orange}{0.753} & \textcolor{orange}{2.20} & \textcolor{orange}{0.663} & \textcolor{orange}{8.00} & 0.746 \\
FireRedTTS-1S-LM & $\checkmark$ & \textcolor{blue}{1.05} & 0.750 & \textcolor{blue}{2.17} & 0.660 & \textcolor{blue}{7.63} & \textcolor{blue}{0.748} \\ \bottomrule
\end{tabular}
\caption{The objective evaluation on Seed-TTS test set. The ranking 1st, 2nd, and 3rd numbers are denoted by ``\textcolor{green}{green}'', ``\textcolor{blue}{blue}'', ``\textcolor{orange}{orange}'' respectively. FireRedTTS-1S-FM and FireRedTTS-1S-LM denote systems with flow-matching-based and language-model-based semantic-to-acoustic decoding approaches, respectively. The closed-source system is denoted by ``*''.}
\label{tab:obj}
\end{table}

\subsection{Subjective Evaluation}
\label{ssec:comos}

The object result demonstrates that FireRedTTS-1S is a high-quality foundation TTS system. However, we also notice that these two metrics, WER and SIM, cannot faithfully represent the performance of a foundation TTS model due to the low coverage of the test set and inaccurate evaluation tools. Moreover, in zero-shot voice cloning, a successful imitation of a voice with high expressiveness but low intelligibility may present a higher WER, which goes against the pursuit of the low WER. Hence, pursuing better WER and SIM on these test sets blindly is not a wise choice during TTS system development. We believe that the result of subjective tests is much more convincing for system evaluation.

In this work, we still employ the evaluation approach, consistency MOS (CoMOS) test, proposed in FireRedTTS-1 \cite{guo2024fireredtts} to measure the consistency between the synthesized speech and prompts (including text and audio). The test set comprises 94 Chinese utterances, covering various emotions and speaking styles. We separate each utterance into two consecutive segments, with the first segment serving as the prompt audio and the second as the synthesis target. During the evaluation, each listener is asked to rate the synthesized audio on a scale of 1 to 5, based on the consistency between synthesized audio and prompts in content, speaking style, and timbre.

We evaluate five groups of audio in this test: the ground-truth audio, the synthesized audio from CosyVoice 2 \cite{du2024cosyvoice}, FireRedTTS-1, FireRedTTS-1S-FM, and FireRedTTS-1S-LM. As presented in Table \ref{tab:exp_content_consistency}, FireRedTTS-1S-FM with the CoMOS of 4.52 and FireRedTTS-1S-LM with the CoMOS of 4.53 achieve impressive synthesis quality, significantly outperforming the previous version with the CoMOS of 4.31 and CosyVoice 2 with the CoMOS of 4.44. It even performs slightly better than the ground-truth audio with the CoMOS of 4.50 in a zero-shot setting. This strongly validates the effectiveness of FireRedTTS-1S as a foundation TTS system.

\begin{table}[htp]
\centering
\begin{tabular}{c|c}
\toprule
\multirow{2}{*}{\textbf{System}} & \multirow{2}{*}{\textbf{CoMOS $\uparrow$}} \\
                                 & \\ \midrule
GroundTruth                      & 4.50 \\
CosyVoice 2.0                    & 4.44 \\
FireRedTTS-1                     & 4.31 \\ \midrule
FireRedTTS-1S-FM                 & 4.52 \\
FireRedTTS-1S-LM                 & \textbf{4.53} \\ \bottomrule
\end{tabular}
\caption{The CoMOS evaluation results.}
\label{tab:exp_content_consistency}
\end{table}

\section{Conclusions}

In this paper, we introduced FireRedTTS-1S, an upgraded streaming foundation text-to-speech system. The system consists of two components: text-to-semantic decoding and semantic-to-acoustic decoding. First, in text-to-semantic decoding, we train a semantic language model to generate semantic speech tokens auto-regressively from the input text and reference audio. Then, in semantic-to-acoustic decoding, we translate semantic tokens into the high-fidelity speech signal in a streaming way. This module is achieved by two approaches: a chunk-wise streamable flow-matching approach and a multi-stream language-model-based approach. After engineering optimization, this system achieves real-time generation with a latency under 150ms. The experimental results demonstrate the synthesis capability of the proposed streaming foundation TTS system. In objective evaluation, it presents a significant improvement over the previous version, achieving comparable intelligibility and speaker similarity to other SOTA TTS systems. Furthermore, the subjective evaluation further highlights its impressive synthesis quality, achieving a subjective score comparable to ground-truth recordings. These results strongly validate FireRedTTS-1S as a high-quality streaming foundation TTS system.

\bibliographystyle{unsrt}
\bibliography{refs}

\end{document}